\documentclass[12pt]{article}
\usepackage[mathscr]{eucal}
\usepackage{amssymb,latexsym}
\usepackage{verbatim}
\usepackage{amsmath}
\usepackage{amsthm}
\usepackage{enumerate}
\usepackage{authblk}
\usepackage{color}
\usepackage{hyperref}

\usepackage[normalem]{ulem}

\newtheorem{thm}{Theorem}
\newtheorem{lem}[thm]{Lemma}
\newtheorem{cor}[thm]{Corollary}

\newcommand\bbR{{\mathbb R}}

\newcommand\e{\epsilon}
\renewcommand\b{\beta}

\newcommand\g{\gamma}

\newcommand\beq{\begin{equation}}
\newcommand\eeq{\end{equation}}
\newcommand\ben{\begin{enumerate}}
\newcommand\een{\end{enumerate}}
\newcommand\bit{\begin{itemize}}
\newcommand\eit{\end{itemize}}

\title{
A note on null distance and causality encoding
}

\author{Gregory J. Galloway\thanks{Research partially supported by the Simons Foundation, Award No. 850541,} 
}

\affil{Department of Mathematics
\\ University of Miami }

\begin{document}
\date{}
\maketitle

\begin{abstract}  

Under natural conditions, the {\it null distance} introduced by Sormani and 
Vega~\cite{SorVega} is a metric space distance function on spacetime, which, in a certain precise sense, can encode the causality of spacetime.  The null distance function requires the choice of a time function. The purpose of this note is to observe that the causality assumptions related to such a choice in results used to establish {\it global} encoding of causality, due Sakovich and Sormani \cite{SakSor} and  to Burtscher and 
García-Heveling \cite{Burt}, can be weakened.

\end{abstract}

\vspace{.2in}


In \cite{SorVega}, C. Sormani and C. Vega introduced a novel distance measure on spacetimes 
(time-oriented  Lorentzian manifolds), called {\it null distance}, which, under reasonable circumstances, has a number of desirable properties. In particular, under certain conditions, it is a metric on spacetime, in the standard sense of metric spaces, and, remarkably, at the same time, encodes the causal structure of spacetime. We will discuss what is meant by the latter shortly.  The construction of such a distance function was motivated in part by the program to study the convergence theory of Lorentzian manifolds.  The null distance function requires the choice of a time function on spacetime. The main purpose of this note is to observe that the causality assumptions related to such a choice, which have been used to establish {\it global} encoding of causality (see \cite{SakSor,Burt}) can  be replaced by an {\it a priori} weaker causality condition.

We begin by fixing some notation and terminology.  By a {\it spacetime\/} we
mean a smooth connected time orientable Lorentzian manifold $(M,g)$, with $\dim M \ge 2$, where $g$ has signature $(-,+ \cdots +)$.
With regard to causal theoretic
notions  such as $I^{\pm}$ (timelike futures and pasts),
$J^{\pm}$ (causal futures and pasts), $D^{\pm}$ (future and past domains of dependence), and $H^{\pm}$ (future and past Cauchy horizons), we by and large follow the conventions in O'Neill \cite{ON}. 

Recall, an edge point of an acausal set $A$ is a point $p$ in the closure of $A$ such that every  neighborhood $U$ of $p$, contains a timelike curve from $I^-(p,U)$ 
to  $I^+(p,U)$ 
that does not meet $A$.  By an {\it acausal $C^0$ spacelike hypersurface} in a spacetime $M$, we mean an  acausal subset $S \subset M$ that does not contain any of its edge points.
It follows  that an acausal $C^0$ spacelike hypersurface is, in fact,
a topological submanifold of $M$ of co-dimension one (see \cite[Proposition~14.25]{ON}).
An acausal $C^0$ spacelike hypersurface $S$ is a {\it Cauchy hypersurface} for $M$ provided every inextendible causal curve in $M$ meets $S$ (and by acausality, does so exactly once). The existence of a Cauchy hypersurface for $M$ is equivalent to $M$ being globally hyperbolic.

A {\it time function} on a spacetime $M$ is a continuous real-valued function $\tau : M \to \bbR$ that is strictly increasing along all future-directed causal curves.   A basic example of a time function is a smooth function on $M$ which has a past pointing timelike gradient.  However, there are types of time functions, such as the so-called cosmological time function \cite{AGH}, that needn't be smooth in general.  The level sets $\tau = \tau_0$ of a time function $\tau$ on $M$ are clearly acausal.  Moreover, it is easy to see, using the continuity of $\tau$, that the level sets do not contain any edge points.  In fact, since the level sets are closed subsets of spacetime (again by continuity of $\tau$), all the level sets are edgeless 
(i.e.\! have no edge points); cf.\! \cite[Corollary 14.26]{ON}.  Thus, the level sets of a time function are edgeless acausal $C^0$ spacelike hypersurfaces. (Such hypersurfaces are sometimes referred to as {\it partial Cauchy surfaces}.) The level sets need not, in general, be connected.

The {\it null distance function} introduced by Sormani and Vega \cite{SorVega} is defined briefly as follows.\footnote{In \cite{SorVega}, the authors define null distance with respect to {\it generalized} time functions, which are not required to be continuous. Many results obtained in \cite{SorVega, Vega, SakSor} apply to such time functions. Here we are only considering continuous time functions.}
Suppose $\tau : M \to \bbR$ is a time function on a spacetime $M$. The {\it null length} of a piecewise causal curve $\b : [a,b] \to M$, with break points $x_i = \beta(s_i)$, $0 \le i \le m$, where each smooth causal segment may be either future pointing or past pointing, is defined by
$$
\hat{L}_\tau(\b) := \sum_{i\, =\, 1}^{m} |\tau(x_{i}) - \tau(x_{i-1})| \,.
$$
The  \emph{null distance function} $\hat{d}_\tau :M \times M \to \bbR$ on $M$ is then defined by taking the inf of the null lengths between any two points $p$ and $q$,
$$
\hat{d}_\tau(p,q) := \inf \{ \hat{L}_\tau (\b) : \b \textrm{\; is piecewise causal from \,} p \textrm{\, to \,} q \} \,.
$$
See e.g.  \cite{SorVega, Vega, SakSor, Burt} for a discussion of fundamental  properties and examples.   

As observed in \cite{SorVega}, the null distance $\hat{d}_\tau :M \times M \to [0, \infty)$ is in general a pseudometric,
but which can  fail to be  definite.  However, there is a natural class of time functions $\tau$ for which the null distance is definite, and hence is a true metric, in the sense of metric spaces.  A time function 
$\tau: M \to \bbR$ is said to be \emph{locally anti-Lipschitz}  provided for every point $p\in M$ there is a neighborhood $U$ of $p$ that has a Riemannian metric with a distance function $d_U: U \times U \to [0,\infty)$  such that for all $x, y\in U$ one has
\beq\
x \le y  \implies \tau(y)-\tau(x) \geq d_U(x,y) \ ,
 \eeq
 where `$x \le y$' means that $y$ is in the causal future of $x$, i.e. $y \in J^+(x)$ (which includes  $y = x$). As shown in \cite{SorVega},  $\hat{d}_\tau:M\times M \to [0, \infty)$ is  definite if and only if $\tau$ is locally anti-Lipschitz.  Moreover, in this case, the metric space topology induced by  $\hat{d}_\tau$ agrees with the manifold topology.  Temporal time functions (smooth functions with past timelike gradient) and  {\it regular} cosmological time functions \cite{AGH} satisfy this locally anti-Lipschitz condtion \cite{SorVega}.  
 
 A remarkable feature of the null distance is that, under favorable circumstances, it encodes the causal structure of spacetime.  We say that $\hat{d}_\tau :M \times M \to [0, \infty)$ encodes causality provided the following holds
 \beq\label{encode}
 p \le q \quad \text{if and only if} \quad \hat{d}_\tau(p,q) = \tau(q) - \tau(p)  \, .
 \eeq
 The `only if' direction follows easily from the definition of null distance.  As shown in \cite{SorVega}, the `if' direction does not hold in general.  However Sormani and Vega \cite[Theorem 3.25]{SorVega} proved that the null distance encodes causality for a large class of generalized Robertson-Walker warped product spacetimes.  
Sakovich and Sormani \cite[Theorem~1.1]{SakSor} proved a local causality encoding result for  time functions $\tau$ that obey the locally anti-Lipschitz condition.  To obtain a global result, they imposed a further condition on the time function. We  paraphrase their result  here.

\begin{thm} [\cite{SakSor}, Theorem 4.1]  \label{SSglobal}
Let $\tau: M \to \bbR$ be a  {\bf proper} locally anti-Lipschitz time function.  Then $\hat{d}_\tau$ globally encodes causality in $M$, i.e. \eqref{encode} holds on $M$.
\end{thm}

\smallskip
Here, $\tau$ being proper means that for all $\tau_1 , \tau_2 \in \tau(M)$,with  $ \tau_1 \le \tau_2$, 
$\tau^{-1}([\tau_1,\tau_2])$ is compact. As observed in \cite{Burt}, the assumption that 
$\tau$ is proper 
implies that its level sets $\tau = \tau_0$ are compact Cauchy hypersurfaces, and hence that $M$ is globally hyperbolic.  It also follows that if all the level sets of $\tau$  are compact Cauchy hypersurfaces then $\tau$ is proper.  Thus, in the statement of Theorem~\ref{SSglobal}, the assumption that $\tau$ is proper can be replaced by the assumption that all level sets of $\tau$ are compact Cauchy hypersurfaces.

From this perspective,  A. Burtscher and L. Garc\'ia-Heveling have recently obtained a generalization of Theorem \ref{SSglobal}.  As described in \cite{Burt}, this is accomplished by combining the approach of Sakovich and Sormani, together with the approach of Burtscher and  Garc\'ia-Heveling in their proof of an earlier version of their global causality encoding result \cite[Theorem 3.3]{Burt}, which required a temporal time function. 

\begin{thm}[\cite{Burt}, Theorem 1.9] \label{BGglobal}
 Let $(M,g)$ be a globally hyperbolic spacetime and $\tau$ a locally anti-Lipschitz time function such that all level sets are Cauchy. Then the null distance encodes causality, i.e., \eqref{encode} holds on $M$.
 \end{thm}

As observed in \cite[Remark 3.6]{Burt}, in the statement of Theorem \ref{BGglobal}, it is sufficient to assume that all level sets of $\tau$ are {\it future Cauchy}.  An acausal $C^0$ spacelike hypersurface $S$ in a spacetime $M$ is future Cauchy if  it has no future Cauchy horizon, i.e. $H^+(S) = \emptyset$. Similarly $S$ is past Cauchy if  $H^-(S) = \emptyset$. Future and past Cauchy hypersurfaces are necessarily connected.
 If S is both future and past Cauchy then it's a Cauchy hypersurface in $M$. 

Here we wish to point out that this future Cauchy condition, under which Theorem~\ref{BGglobal} holds, can be replaced by a simple causal theoretic condition which {\it in general} is much weaker.
An acausal $C^0$ spacelike hypersurface is said to be {\it future causally complete} provided for all points $p \in J^+(S)$, the set $J^-(p) \cap S$ has compact closure~in~S. (Past causal completeness is defined in a time-dual manner.)  As a very simple example, consider Minkowski space N, with standard coordinates $(t, x^1, \cdots, x^n)$, and let $M=N \setminus\{(1, 0,\cdots, 0)\}$.  The time slices $t = t_0 < 1$ are future causally complete, but not future Cauchy.  Similarly, the spacelike hypersurfaces $t = \sqrt{a^2 + \sum_i (x^i)^2}$, $a \in (0,1)$ are future causally complete, but neither future nor past Cauchy.  

Note that all compact acausal $C^0$ hypersurfaces, as well as all future Cauchy hypersurfaces, are future causally complete. (The latter follows from e.g.\! \cite[Prop.~5.20]{Pen}.  In fact, if $S$ is future Cauchy then $J^-(p) \cap S$ is compact for all $p \in J^+(S)$.)  Future causal completeness was introduced in \cite{GGcamb}, and used, for example,  to extend Hawking's cosmological singularity theorem to the nonspatially closed case in spacetimes not assumed to be globally hyperbolic.  Further applications have been considered in e.g. \cite{GGSome, Grant, horo1}.

\smallskip
We have the following refinement of Theorem \ref{BGglobal}.

\begin{thm} \label{Gglobal}
 Let $\tau$ be a locally anti-Lipschitz time function on a spacetime $(M,g)$ such that all the level sets are future causally complete. Then the null distance encodes causality, i.e.\! \eqref{encode} holds on $M$. Furthermore, $(M,g)$ is globally hyperbolic.
\end{thm}

The theorem is an immediate consequence of \cite[Theorem 1.9, Remark 3.6]{Burt} and the following  observation.

\begin{lem} \label{fcc}
 Let $\tau$ be a time function on a spacetime $(M,g)$ such that all the level sets are future causally complete. Then all the level sets are future Cauchy. In particular, $(M,g)$ is globally hyperbolic.
\end{lem}

\proof  The lemma follows from a much more general result \cite[Theorem 2.2]{Horta} concerning spacetime foliations. Here we give a simple direct proof applicable to spacetimes with time functions.  Let $S_a$ denote the level set $\tau = a$. It suffices to show that for each $a \in \tau(M)$, $H^+(S_a) = \emptyset$.  Suppose to the contrary, for some $a$ there exists a point $q \in H^+(S_a)$.  Then $q$ is the future endpoint of a past directed null geodesic $\eta$ contained in $H^+(S_a)$, which is either past inextendible or has a past end point on an edge point of $S_a$ (\cite[Theorem 5.12]{Pen}).  But $S_a$ is edgless. Hence $\eta: [0,c) \to M$, $s \to \eta(s)$, is past inextendible in $H^+(S_a)$.  Let $\{s_k\}$ be an increasing sequence of numbers such that $s_k \to c$, and let $q_k = \eta(s_k)$.  Then $\{\tau(q_k)\}$ is a decreasing sequence of numbers; set  $b = \inf \{\tau(q_k)\} \ge a$.  

Consider the case $b > a$. Fix a smooth, past directed, timelike vector field $T$ on $M$. Let $\g_k$ be an integral curve of $T$, starting at $q_k$.  $\g_k$ meets $S_a$ at a point $p_k$, say (\cite[Lemma 14.51]{ON}). Since $\tau(p_k) = a < b < \tau(q_k)$, continuity of $\tau$ implies that $\g_k$ meets $S_b$ at a point $r_k$, say.  By the future causal completeness of $S_b$, $J^-(q_1) \cap J^+(S_b)$ has compact closure in $S_b$.  Hence, by passing to a subsequence if necessary, we may assume that $r_k$ converges to a point $r \in S_b$.  

Let $U$ be a small neighborhood of $r$ in $S_b$
 with compact closure.  
 Using the flow of the vector field $T$ near $\overline{U}$, we obtain a small  compact spacetime neighborhood $V \cong \overline{U} \times [-\e,\e]$ of $r$, which is  contained in $D(S_b) = 
D^+(S_b) \cup D^-(S_b)$. Let $A$ be the set of all points $x \in M$ such that $x$ lies on an integral curve of $T$ that passes through $\overline{U}$.  Then, by continuity of $\tau$, there exists $\delta >0$ such that
$$
\{x \in A: b-\delta < \tau(x) < b+ \delta \}  \subset V \, .
$$
It follows that for $k$ sufficiently large, $q_k \in V \cap J^+(S_b) \subset D^+(S_b)$.  But this means that the null geodesic $\eta$ must extend to $S_b$, which is a contradiction.  Hence, 
in this case, $H^+(S_a)~=~\emptyset$.  

The argument for the case $b = a$ is essentially the same. The future causal completeness of $S_a$, implies that the $p_k's$ subconverge to a point $p \in S_a$.  Now apply the same argument around the point $p$ as was done around the point $r$ above.  Thus, in either case, we conclude that $S_a$ is future Cauchy.
That $(M,g)$ is globally hyperbolic follows from the fact that $I^+(S_a) = {\rm int}\, D^+(S_a)$ is globally hyperbolic for all $a \in \tau(M)$.
\qed

\smallskip
Theorem \ref{Gglobal}  immediate implies the following refinement of Theorem \ref{SSglobal}.

\begin{cor}\label{Gglobal2}
 Let $\tau$ be a locally anti-Lipschitz time function on a spacetime $(M,g)$ such that all level sets are compact. Then the null distance encodes causality, i.e., \eqref{encode} holds on $M$. Furthermore, $(M,g)$ is globally hyperbolic.
\end{cor}

\smallskip
\noindent
{\it Remark.}  In this case  the level sets are Cauchy hypersurfaces. This can be seen, for example,  by applying Lemma \ref{fcc} and its time dual to each level  surface.



\begin{thebibliography}{10}

\bibitem{AGH}
L.~Andersson, G.~J. Galloway, and R.~Howard, \emph{The cosmological time
  function}, Classical Quantum Gravity \textbf{15} (1998), no.~2, 309--322.

\bibitem{Burt}
A.~Burtscher and L.~García-Heveling, \emph{Global hyperbolicity through the
  eyes of the null distance}, 2022, arXiv:2209.15610 [math.DG].

\bibitem{GGcamb}
G.~J. Galloway, \emph{Curvature, causality and completeness in space-times with
  causally complete spacelike slices}, Math. Proc. Cambridge Philos. Soc.
  \textbf{99} (1986), no.~2, 367--375.

\bibitem{GGSome}
\bysame, \emph{Some connections between global hyperbolicity and geodesic
  completeness}, J. Geom. Phys. \textbf{6} (1989), no.~1, 127--141.

\bibitem{horo1}
G.~J. Galloway and C.~Vega, \emph{Achronal limits, lorentzian spheres, and
  splitting}, Ann. Henri Poincar\'e \textbf{15} (2014), no.~11, 2241--2279.

\bibitem{Horta}
A.~Horta, \emph{Studies in {L}orentzian geometry and mathematical relativity},
  ProQuest LLC, Ann Arbor, MI, 1993, Thesis (Ph.D.)--University of Miami.

\bibitem{ON}
B.~O'Neill, \emph{Semi-{R}iemannian geometry}, Pure and Applied Mathematics,
  vol. 103, Academic Press Inc. [Harcourt Brace Jovanovich Publishers], New
  York, 1983.

\bibitem{Pen}
R.~Penrose, \emph{Techniques of differential topology in relativity}, Society
  for Industrial and Applied Mathematics, Philadelphia, Pa., 1972, Conference
  Board of the Mathematical Sciences Regional Conference Series in Applied
  Mathematics, No.~7.

\bibitem{SakSor}
A.~Sakovich and C.~Sormani, \emph{{The null distance encodes causality}},
  Journal of Mathematical Physics \textbf{64} (2023), no.~1, 012502.

\bibitem{SorVega}
C.~Sormani and C.~Vega, \emph{Null distance on a spacetime}, Classical Quantum
  Gravity \textbf{33}, no.~8, 085001.

\bibitem{Grant}
J.-H. Treude and J.~D.~E. Grant, \emph{Volume comparison for hypersurfaces in
  {L}orentzian manifolds and singularity theorems}, Ann. Global Anal. Geom.
  \textbf{43} (2013), no.~3, 233--251.

\bibitem{Vega}
C.~Vega, \emph{Spacetime distances: an exploration}, 2021, arXiv:2103.01191
  [gr-qc].

\end{thebibliography}

\providecommand{\bysame}{\leavevmode\hbox to3em{\hrulefill}\thinspace}
\providecommand{\MR}{\relax\ifhmode\unskip\space\fi MR }
\providecommand{\MRhref}[2]{%
  \href{http://www.ams.org/mathscinet-getitem?mr=#1}{#2}
}
\providecommand{\href}[2]{#2}

\end{document}